\documentclass[%
 reprint,
 amsmath,amssymb,
 aps,
]{revtex4-2}

\usepackage{graphicx}
\usepackage{dcolumn}
\usepackage{bm}

\begin{document}

\preprint{APS/123-QED}

\title{Volatility-inspired $\sigma$-LSTM cell}

\author{German Rodikov}
 \altaffiliation{Scuola Normale Superiore, Pisa, Italy}

\author{Nino Antulov-Fantulin}
\altaffiliation{ETH Zürich, Switzerland \\ Aisot Technologies AG, Zürich, Switzerland}

\date{\today}

\begin{abstract}
Volatility models of price fluctuations are well studied in the econometrics literature, with more than 50 years of theoretical and empirical findings. The recent advancements in neural networks (NN) in the deep learning field have naturally offered novel econometric modeling tools. However, there is still a lack of explainability and stylized knowledge about volatility modeling with neural networks; the use of stylized facts could help improve the performance of the NN for the volatility prediction task. In this paper, we investigate how the knowledge about the "physics" of the volatility process can be used as an inductive bias to design or constrain a cell state of long short-term memory (LSTM) for volatility forecasting. We introduce a new type of $\sigma$-LSTM cell with a stochastic processing layer, design its learning mechanism and show good out-of-sample forecasting performance. 
\end{abstract}

\maketitle


\section{Introduction}

The structure of noise or errors in regression models is usually subtle and taken as an ansatz to use different mathematical frameworks. E.g. in the case of linear regression models $ \mathbf{y} = \mathbf{X} \beta + \epsilon$, where $\mathbf{y} \in \mathbb {R} ^{n}$ represent response variable, unobservable parameters $\beta \in \mathbb {R} ^{K}$ and non-random explanatory variable $\mathbf{X} \in \mathbb {R} ^{n\times K}$, while $\epsilon$ represents the noise or error. 
When the errors $\epsilon_{i}$ are homoscedastic i.e. $Var[\epsilon_i]=\sigma^2$ and serially uncorrelated i.e. $Cov[\epsilon_i, \epsilon_j]=0$ by the Gauss–Markov theorem~\cite{huang1970regression} the ordinary least squares is having the lowest sampling variance within the class of linear unbiased estimators. In econometrics, special interpretation and attention are given to the error structure~\cite{tsay2005analysis}. 

In this paper, we focus on the problem of volatility of asset returns, which are well known to be heteroscedastic in nature. Volatility is associated with the risk and amplitude of price fluctuations. 

Some models characterize the volatility from a conditional process perspective, for example, the autoregressive conditional heteroskedasticity (ARCH) model~\cite{Engle_1982} and the generalized autoregressive conditional heteroskedasticity (GARCH) model~\cite{Bollerslev_1986}. The idea of the conditional process approach is the possibility of using a conditional variance that varies over time while the unconditional variance remains relatively constant. 

On the other hand, researchers have recently focused on using realized volatility (RV) to build forecasting models due to the wide availability of high-frequency financial data. For example, the Heterogeneous Autoregression (HAR-RV) model~\cite{corsi2009simple} is widely used in the literature due to consistently good predictive performance and simple methods to estimate it. 

Recently, different recurrent NN units like LSTM~\cite{LSTM_hochreiter1997}, GRU~\cite{GRU_cho2014}, SRU~\cite{oliva2017statistical} have demonstrated high performance on forecasting tasks for time-series data. More specifically, LSTM is the particular architecture that design improves the model's performance overall and especially in the volatility prediction task~\cite{bucci2020realized, rodikov_antulov2022}. 

Gated recurrent units (GRU)~\cite{GRU_cho2014} is an enhanced LSTM architecture that improves the fitting process by eliminating the cell state. In addition, the Statistical Recurrent Unit (SRU)~\cite{oliva2017statistical} was introduced, which can infer long-term dependencies from data by using simple moving averages of summary statistics and has multiple proxies of the past with simple linear combinations.

Our work investigated and analyzed how NN can learn to capture the temporal structure of realized volatility. We are interested in how we could add the structure of long and short-term volatility effects to the LSTM. We introduce a modified LSTM cell that we call $\sigma$-LSTM to match these needs. For this reason, we extend the equation system of the LSTM cell, which reflects the inductive bias of GARCH-like structure and HAR-RV effects of volatility, which allows easier learning of volatility by maximum likelihood estimation.

Recently, several studies~\cite{nguyen2020recurrent,nguyen2022statistical} have explored the heteroskedasticity of returns with recurrent NN architectures, but not by means of modified LSTM cell.
In~\cite{nguyen2020recurrent}, authors have proposed the RECH model, where $\omega$-constant of the GARCH process is modeled by a particular RNN model. 

In~\cite{nguyen2022statistical}, authors have proposed the combination of the Stochastic Volatility (SV) model and Statistical Recurrent Unit (SRU). The idea is that the SRU captures the long-term memory effects and auto-dependence of the volatility. However, the SRU is modeling the deterministic dynamics of the hidden states in the SR-SV model. 

We investigated the original Long short-term memory cell, proposed $\sigma$-LSTM with a particular loss function for realized volatility forecasting tasks, and compared the predictive ability with widely used HAR-RV, GARCH(1,1) models.

The remaining paper is organized as follows. Section II provides mathematical motivation and a formal definition of $\sigma$-LSTM cell. Section III describes how the experiment and its results. Finally, in Section IV, we provide a conclusion. 

\section{Methodology}

One important class of econometric models are GARCH family models~\cite{Engle_1982, Bollerslev_1986}:

\begin{equation} \label{GARCH}
       r_{t}=\mu_t+\sigma_{t} \varepsilon_{t}, \text{ } \sigma_{t}\varepsilon_{t} \sim \mathcal{N}(0,\sigma_{t}^2)
\end{equation}

where the conditional variance~\cite{Bollerslev_1986, Engle_1982} has the autoregressive structure:

\begin{equation} \label{GARCH1}
\sigma_{\mathrm{t}}^{2}=\omega+\alpha * \mathrm{r}_{\mathrm{t}-1}^{2}+\beta * \sigma_{\mathrm{t}-1}^{2}.
\end{equation}

Number extensions with different functional dependence (see Table~\ref{tab:GARCH}) have been proposed like eGARCH~\cite{nelson1991conditional}, cGARCH~\cite{cGARCH}, TGARCH~\cite{TARCH}, GJR-GARCH~\cite{glosten1993relation} and others.

\vspace{0.1cm}
\begin{table}[ht]
\centering
\caption{GARCH family}
\label{tab:GARCH}       
\begin{tabular}{c} 
\hline
\hline \\
eGARCH \\
$\ln({\sigma}_t^2) = \omega + \alpha \left[ \left| \frac{{\varepsilon}_{t-1}}{{\sigma}_{t-1}} \right| - E\left| \frac{{\varepsilon}_{t-1}}{{\sigma}_{t-1}} \right|  \right] + \delta \frac{{\varepsilon}_{t-1}}{{\sigma}_{t-1}} + \beta \ln ({\sigma}_{t-1}^2)$\\ \\
\hline \\
cGARCH \\
${\sigma}_t^2 = q_t + \alpha ({\varepsilon}^2_{t-1} - q_{t-1})+ \beta ({\sigma}_{t-1}^2 - q_{t-1})$ \\
        $q_t = \omega  + \rho q_{t-1} + \theta ({\varepsilon}^2_{t-1} - {\sigma}^2_{t-1})$\\ \\
\hline \\
GJR-GARCH \\
${\sigma}_t^2=\omega+\left(\alpha+\gamma I_{t-1}\right) \varepsilon_{t-1}^{2}+\beta \sigma_{t-1}^{2}$ \\
       $I_{t-1} \{\begin{array}{ll} 
      0 & \text { if } r_{t-1} \geq \mu \\ 
      1 & \text { if } r_{t-1}<\mu\end{array}$\\ \\
\hline \\
TGARCH \\

$\sigma_{t}=\omega+\alpha \varepsilon_{t-1}+\beta \sigma_{t-1}+\phi \varepsilon_{t-1} 1_{\left[\varepsilon_{t-1}<0\right]}$ \\ \\
\hline
\hline
\end{tabular}
\end{table} 
\vspace{0.1cm}

The heterogeneous Autoregression Realized Volatility (HAR-RV)  model introduced by~\cite{corsi2009simple} assumes that agents' behavior in financial markets, which differ in their perception of volatility depending on their investment horizons and are divided into short-term, medium-term, and long-term. Heterogeneous structures in financial markets are based on the heterogeneous market hypothesis presented by~\cite{muller1993fractals}. Participants' decisions refer to different time horizons that perceive and respond to different types of volatility. A memory of each component decreases with a particular time constant.

The HAR-RV model is an additive cascade of partial volatilities generated at different time horizons that follows an autoregressive process~\cite{corsi2009simple}. The HAR-RV approach is one more stable and accurate estimate for Realized Volatility~\cite{corsi2012har} at the 3 different horizons, where $R V_{t}^{(d)}, R V_{t}^{(w)}$, and $R V_{t}^{(m)}$ are respectively the daily, weekly, and monthly observed realized volatilities.

\vspace{0.1cm}
\begin{equation} 
    \begin{cases}
        \tilde{\sigma}_{t+1 m}^{(m)}=c^{(m)}+\phi^{(m)} R V_{t}^{(m)}+\tilde{\omega}_{t+1 m}^{(m)} \\
        \tilde{\sigma}_{t+1 w}^{(w)}=c^{(w)}+\phi^{(w)} R V_{t}^{(w)}+\gamma^{(w)} \mathbb{E}_{t}\left[\tilde{\sigma}_{t+1 m}^{(m)}\right]+\tilde{\omega}_{t+1 w^{\prime}}^{(w)} \\
        \tilde{\sigma}_{t+1 d}^{(d)}=c^{(d)}+\phi^{(d)} R V_{t}^{(d)}+\gamma^{(d)} \mathbb{E}_{t}\left[\tilde{\sigma}_{t+1 w}^{(w)}\right]+\tilde{\omega}_{t+1 d}^{(d)}
    \end{cases}
\end{equation}
\vspace{0.1cm}
where $c^{(m)}$ - the constant and $\tilde{\omega}_{t+1 m}^{(m)}$ is an innovation that is simultaneously and consistently independent with a mean zero for monthly aggregation, and $\phi$ represents the wight in a particular cascade. 

Motivated by GARCH structure and long and short-term volatility effect of HAR-RV model, we propose to model the return assets
$\mathrm{r}_{\mathrm{t}}=\phi(\mathrm{r}_{\mathrm{t}-1},..., \mathrm{r}_{\mathrm{t}-p})$, where $\phi(.)$ is a differentiable non-linear function. In particular a modified long short-term memory (LSTM) cell~\cite{LSTM_hochreiter1997} that should capture long and short-term volatility.
The inputs to out modified LSTM cell $x_t=r_t$ are directly returns and the outputs are $\hat{r}_{t}$ and $\hat{\sigma}^2_{t}$. The cell has directly the hidden representation $h_{t}$ for short-term memory and long-term $C_{t}$ volatility memory component. The updates rules of $\sigma-$LSTM are the following:

\begin{equation} \label{LSTM_GP}
f_{t}=\sigma\left(W_{f} \cdot\left[h_{t-1}, x_{t}\right]+b_{f}\right),
\end{equation}
\begin{equation} 
i_{t} =\sigma\left(W_{i} \cdot\left[h_{t-1}, x_{t}\right]+b_{i}\right),
\end{equation}
\begin{equation} 
\tilde{C}_{t} =\tanh \left(W_{C}
 \cdot\left[h_{t-1},x_{t}\right]+b_{C}\right),
\end{equation}
\begin{equation} 
C_{t}=f_{t} * C_{t-1}+i_{t} * \tilde{C}_{t},
\end{equation}
\begin{equation} 
o_{t} ={\mathcal {N}}(0 ,W_{o}[{C}_{t}^{2}]),
\end{equation}
\begin{equation} 
h_{t} = o_{t} * \phi \left(C_{t}\right).
\end{equation}
Finally, the output return $\hat{r}_{t}$ and estimated volatility $\hat{\sigma}_{t}$ is:
\begin{equation} 
\hat{r}_{t} = W_{h}  \cdot h_{t}
\end{equation}
\begin{equation} 
\hat{\sigma}^2_{t} = {\langle C_{t} \rangle}^2,
\end{equation}
where $\langle . \rangle$ is the mean operator and both $\hat{r}_{t}$ and $\hat{\sigma}^2_{t}$ are scalar values.

We implement the custom loss function as the likelihood of observed returns with estimated volatilities.

\begin{equation} \label{Custom_loss}
\mathcal{L} = \sum_{t=1}^{m}\left[-\ln \left(\hat{\sigma}_{t}^{2}\right)-\frac{r_{t}^{2}}{\hat{\sigma}_{t}^{2}}\right].
\end{equation}

\section{Experiments \& Results}

\begin{table}[h]
\caption{\label{data}Description of the data }
\begin{ruledtabular}
\begin{tabular}{cccc}
Type of asset & Name &     Price points  & RV points  \\
\hline
Index & S\&P500 & 2 821 368 & 3803 \\
Stock & Apple Inc.  & 2 466 466 & 3803 \\
Cryptocurrency & Bitcoin-USD   & 3 613 769 & 3375  \\
\end{tabular}
\end{ruledtabular}
\end{table}

This study investigates how proposed $\sigma$-LSTM could estimate and predict realized volatility on different market structures, particularly stocks, indexes, and cryptocurrency data. We consider Apple inc. stock, the S\&P~500 index, and Bitcoin-USD. 

We calculate RV based on minutes-based price observations for daily aggregation. Returns are calculated on the daily close price. As a best practice, we divided the dataset into three parts: training, validation, and test. The validation and the test sample are equivalent to 200 points.

Mean Squared Error (MSE) measures averaged squared difference values between the predictions and the target. The power of 2 in this metric prevents neutralizing positive and negative deviations, which minimizes the distance between actual and calculated values. Root Mean Squared Error (RMSE) is the square root of MSE. The square root is introduced to scale error is the same as the target scale.

To find the best configuration of NN is necessary to conduct multiple experiments with different hyperparameters~\cite{panchal2010searching}. We have results for training launches and results for the validation dataset; the next step is to select promising hyperparameters RMSE metrics appropriately. Standardization is highly recommended before training RNNs and can improve the efficiency of training models. We normalized input data from 0 to 1 by min-max scale.

In our study, we ask the following questions. First, how could the proposed $\sigma$-LSTM cell with GARCH-like structure and long and short-term volatility effect of the HAR-RV model capture the long and short-term volatility effects? 

We performed standard accuracy measures for the one-step-ahead prediction using RMSE metrics for 200 data points, Table ~\ref{table:result}. As a result, $\sigma$-LSTM shows the best performance for RMSE for the out-of-sample result for the S\&P~500 index and Apple Inc. stock data sets.

\begin{table}[b]
\caption{\label{table:result}%
S\&P~500, Apple Inc. stock and Bitcoin-USD out-of-sample tests of forecasting accuracy }
\begin{ruledtabular}
\begin{tabular}{ccc}
Data Set & Model  &  RMSE\\
\hline
S\&P~500 Index    &GARCH (1,1)             &  0.00405  \\
&HAR-RV                  &  0.00359   \\
&LSTM                    &  0.00805   \\
&$\sigma$-LSTM             &  0,00351 \\
\hline
Apple Inc. stock &GARCH (1,1)     &  0.00648    \\
&HAR-RV          &  0.00561   \\
&LSTM            &  0.00752  \\
&$\sigma$-LSTM        & 0.00560 \\
\hline
Bitcoin-USD &GARCH (1,1)     &  0.01641    \\
&HAR-RV          &  0.01537   \\
&LSTM            &  0.02286  \\
&$\sigma$-LSTM      &  0.01542   \\
\end{tabular}
\end{ruledtabular}
\end{table}

However, it should be noted that in the case of cryptocurrency, the prediction error of HAR-RV was at the same level as $\sigma$-LSTM. In our experiment, $C_{t}$ of the original LSTM does not provide sufficient results.

\section{Conclusion}
This work introduces a special $\sigma$-LSTM cell to investigate whether the use of stylized facts or "physics-informed" inductive bias~\cite{karniadakis2021physics} i.e., GARCH and HAR-RV volatility structure could help to improve the performance of the NN for the volatility prediction task. We do not use the Recurrent LSTM unit as a black box but rather design a sub-component to represent a long-short volatility memory and a stochastic part.

We add particular loss functions for the $\sigma$-LSTM. As a result, we show that $\sigma$-LSTM could outperform well-known models in this field, such as a strong baseline HAR-RV and regular LSTM cell. We will investigate more advanced loss functions in future work that could allow faster learning convergence and accuracy.

\bibliography{my_bib}

\begin{thebibliography}{20}%
\makeatletter
\providecommand \@ifxundefined [1]{%
 \@ifx{#1\undefined}
}%
\providecommand \@ifnum [1]{%
 \ifnum #1\expandafter \@firstoftwo
 \else \expandafter \@secondoftwo
 \fi
}%
\providecommand \@ifx [1]{%
 \ifx #1\expandafter \@firstoftwo
 \else \expandafter \@secondoftwo
 \fi
}%
\providecommand \natexlab [1]{#1}%
\providecommand \enquote  [1]{``#1''}%
\providecommand \bibnamefont  [1]{#1}%
\providecommand \bibfnamefont [1]{#1}%
\providecommand \citenamefont [1]{#1}%
\providecommand \href@noop [0]{\@secondoftwo}%
\providecommand \href [0]{\begingroup \@sanitize@url \@href}%
\providecommand \@href[1]{\@@startlink{#1}\@@href}%
\providecommand \@@href[1]{\endgroup#1\@@endlink}%
\providecommand \@sanitize@url [0]{\catcode `\\12\catcode `\$12\catcode
  `\&12\catcode `\#12\catcode `\^12\catcode `\_12\catcode `\%12\relax}%
\providecommand \@@startlink[1]{}%
\providecommand \@@endlink[0]{}%
\providecommand \url  [0]{\begingroup\@sanitize@url \@url }%
\providecommand \@url [1]{\endgroup\@href {#1}{\urlprefix }}%
\providecommand \urlprefix  [0]{URL }%
\providecommand \Eprint [0]{\href }%
\providecommand \doibase [0]{https://doi.org/}%
\providecommand \selectlanguage [0]{\@gobble}%
\providecommand \bibinfo  [0]{\@secondoftwo}%
\providecommand \bibfield  [0]{\@secondoftwo}%
\providecommand \translation [1]{[#1]}%
\providecommand \BibitemOpen [0]{}%
\providecommand \bibitemStop [0]{}%
\providecommand \bibitemNoStop [0]{.\EOS\space}%
\providecommand \EOS [0]{\spacefactor3000\relax}%
\providecommand \BibitemShut  [1]{\csname bibitem#1\endcsname}%
\let\auto@bib@innerbib\@empty
\bibitem [{\citenamefont {Huang}(1970)}]{huang1970regression}%
  \BibitemOpen
  \bibfield  {author} {\bibinfo {author} {\bibfnamefont {D.~S.}\ \bibnamefont
  {Huang}},\ }\href@noop {} {\emph {\bibinfo {title} {Regression and
  econometric methods}}},\ \bibinfo {number} {QA 278.2. H82}\ (\bibinfo {year}
  {1970})\BibitemShut {NoStop}%
\bibitem [{\citenamefont {Tsay}(2005)}]{tsay2005analysis}%
  \BibitemOpen
  \bibfield  {author} {\bibinfo {author} {\bibfnamefont {R.~S.}\ \bibnamefont
  {Tsay}},\ }\href@noop {} {\emph {\bibinfo {title} {Analysis of financial time
  series}}}\ (\bibinfo  {publisher} {John wiley \& sons},\ \bibinfo {year}
  {2005})\BibitemShut {NoStop}%
\bibitem [{\citenamefont {Engle}(1982)}]{Engle_1982}%
  \BibitemOpen
  \bibfield  {author} {\bibinfo {author} {\bibfnamefont {R.~F.}\ \bibnamefont
  {Engle}},\ }\bibfield  {title} {\bibinfo {title} {Autoregressive conditional
  heteroscedasticity with estimates of the variance of united kingdom
  inflation},\ }\href@noop {} {\bibfield  {journal} {\bibinfo  {journal}
  {Econometrica: Journal of the econometric society}\ ,\ \bibinfo {pages}
  {987}} (\bibinfo {year} {1982})}\BibitemShut {NoStop}%
\bibitem [{\citenamefont {Bollerslev}(1986)}]{Bollerslev_1986}%
  \BibitemOpen
  \bibfield  {author} {\bibinfo {author} {\bibfnamefont {T.}~\bibnamefont
  {Bollerslev}},\ }\bibfield  {title} {\bibinfo {title} {Generalized
  autoregressive conditional heteroskedasticity},\ }\href@noop {} {\bibfield
  {journal} {\bibinfo  {journal} {Journal of econometrics}\ }\textbf {\bibinfo
  {volume} {31}},\ \bibinfo {pages} {307} (\bibinfo {year} {1986})}\BibitemShut
  {NoStop}%
\bibitem [{\citenamefont {Corsi}(2009)}]{corsi2009simple}%
  \BibitemOpen
  \bibfield  {author} {\bibinfo {author} {\bibfnamefont {F.}~\bibnamefont
  {Corsi}},\ }\bibfield  {title} {\bibinfo {title} {A simple approximate
  long-memory model of realized volatility},\ }\href@noop {} {\bibfield
  {journal} {\bibinfo  {journal} {Journal of Financial Econometrics}\ }\textbf
  {\bibinfo {volume} {7}},\ \bibinfo {pages} {174} (\bibinfo {year}
  {2009})}\BibitemShut {NoStop}%
\bibitem [{\citenamefont {Hochreiter}\ and\ \citenamefont
  {Schmidhuber}(1997)}]{LSTM_hochreiter1997}%
  \BibitemOpen
  \bibfield  {author} {\bibinfo {author} {\bibfnamefont {S.}~\bibnamefont
  {Hochreiter}}\ and\ \bibinfo {author} {\bibfnamefont {J.}~\bibnamefont
  {Schmidhuber}},\ }\bibfield  {title} {\bibinfo {title} {Long short-term
  memory},\ }\href@noop {} {\bibfield  {journal} {\bibinfo  {journal} {Neural
  computation}\ }\textbf {\bibinfo {volume} {9}},\ \bibinfo {pages} {1735}
  (\bibinfo {year} {1997})}\BibitemShut {NoStop}%
\bibitem [{\citenamefont {Cho}\ \emph {et~al.}(2014)\citenamefont {Cho},
  \citenamefont {Van~Merri{\"e}nboer}, \citenamefont {Gulcehre}, \citenamefont
  {Bahdanau}, \citenamefont {Bougares}, \citenamefont {Schwenk},\ and\
  \citenamefont {Bengio}}]{GRU_cho2014}%
  \BibitemOpen
  \bibfield  {author} {\bibinfo {author} {\bibfnamefont {K.}~\bibnamefont
  {Cho}}, \bibinfo {author} {\bibfnamefont {B.}~\bibnamefont
  {Van~Merri{\"e}nboer}}, \bibinfo {author} {\bibfnamefont {C.}~\bibnamefont
  {Gulcehre}}, \bibinfo {author} {\bibfnamefont {D.}~\bibnamefont {Bahdanau}},
  \bibinfo {author} {\bibfnamefont {F.}~\bibnamefont {Bougares}}, \bibinfo
  {author} {\bibfnamefont {H.}~\bibnamefont {Schwenk}},\ and\ \bibinfo {author}
  {\bibfnamefont {Y.}~\bibnamefont {Bengio}},\ }\bibfield  {title} {\bibinfo
  {title} {Learning phrase representations using rnn encoder-decoder for
  statistical machine translation},\ }\href@noop {} {\bibfield  {journal}
  {\bibinfo  {journal} {arXiv preprint arXiv:1406.1078}\ } (\bibinfo {year}
  {2014})}\BibitemShut {NoStop}%
\bibitem [{\citenamefont {Oliva}\ \emph {et~al.}(2017)\citenamefont {Oliva},
  \citenamefont {P{\'o}czos},\ and\ \citenamefont
  {Schneider}}]{oliva2017statistical}%
  \BibitemOpen
  \bibfield  {author} {\bibinfo {author} {\bibfnamefont {J.~B.}\ \bibnamefont
  {Oliva}}, \bibinfo {author} {\bibfnamefont {B.}~\bibnamefont {P{\'o}czos}},\
  and\ \bibinfo {author} {\bibfnamefont {J.}~\bibnamefont {Schneider}},\
  }\bibfield  {title} {\bibinfo {title} {The statistical recurrent unit},\ }in\
  \href@noop {} {\emph {\bibinfo {booktitle} {International Conference on
  Machine Learning}}}\ (\bibinfo {organization} {PMLR},\ \bibinfo {year}
  {2017})\ pp.\ \bibinfo {pages} {2671--2680}\BibitemShut {NoStop}%
\bibitem [{\citenamefont {Bucci}(2020)}]{bucci2020realized}%
  \BibitemOpen
  \bibfield  {author} {\bibinfo {author} {\bibfnamefont {A.}~\bibnamefont
  {Bucci}},\ }\bibfield  {title} {\bibinfo {title} {Realized volatility
  forecasting with neural networks},\ }\href@noop {} {\bibfield  {journal}
  {\bibinfo  {journal} {Journal of Financial Econometrics}\ }\textbf {\bibinfo
  {volume} {18}},\ \bibinfo {pages} {502} (\bibinfo {year} {2020})}\BibitemShut
  {NoStop}%
\bibitem [{\citenamefont {Rodikov}\ and\ \citenamefont
  {Antulov-Fantulin}(2022)}]{rodikov_antulov2022}%
  \BibitemOpen
  \bibfield  {author} {\bibinfo {author} {\bibfnamefont {G.}~\bibnamefont
  {Rodikov}}\ and\ \bibinfo {author} {\bibfnamefont {N.}~\bibnamefont
  {Antulov-Fantulin}},\ }\bibfield  {title} {\bibinfo {title} {Can lstm
  outperform volatility-econometric models?},\ }\href@noop {} {\bibfield
  {journal} {\bibinfo  {journal} {arXiv preprint arXiv:2202.11581}\ } (\bibinfo
  {year} {2022})}\BibitemShut {NoStop}%
\bibitem [{\citenamefont {Nguyen}\ \emph {et~al.}(2020)\citenamefont {Nguyen},
  \citenamefont {Tran},\ and\ \citenamefont {Kohn}}]{nguyen2020recurrent}%
  \BibitemOpen
  \bibfield  {author} {\bibinfo {author} {\bibfnamefont {T.-N.}\ \bibnamefont
  {Nguyen}}, \bibinfo {author} {\bibfnamefont {M.-N.}\ \bibnamefont {Tran}},\
  and\ \bibinfo {author} {\bibfnamefont {R.}~\bibnamefont {Kohn}},\ }\bibfield
  {title} {\bibinfo {title} {Recurrent conditional heteroskedasticity},\
  }\href@noop {} {\bibfield  {journal} {\bibinfo  {journal} {arXiv preprint
  arXiv:2010.13061}\ } (\bibinfo {year} {2020})}\BibitemShut {NoStop}%
\bibitem [{\citenamefont {Nguyen}\ \emph {et~al.}(2022)\citenamefont {Nguyen},
  \citenamefont {Tran}, \citenamefont {Gunawan},\ and\ \citenamefont
  {Kohn}}]{nguyen2022statistical}%
  \BibitemOpen
  \bibfield  {author} {\bibinfo {author} {\bibfnamefont {T.-N.}\ \bibnamefont
  {Nguyen}}, \bibinfo {author} {\bibfnamefont {M.-N.}\ \bibnamefont {Tran}},
  \bibinfo {author} {\bibfnamefont {D.}~\bibnamefont {Gunawan}},\ and\ \bibinfo
  {author} {\bibfnamefont {R.}~\bibnamefont {Kohn}},\ }\bibfield  {title}
  {\bibinfo {title} {A statistical recurrent stochastic volatility model for
  stock markets},\ }\href@noop {} {\bibfield  {journal} {\bibinfo  {journal}
  {Journal of Business \& Economic Statistics}\ ,\ \bibinfo {pages} {1}}
  (\bibinfo {year} {2022})}\BibitemShut {NoStop}%
\bibitem [{\citenamefont {Nelson}(1991)}]{nelson1991conditional}%
  \BibitemOpen
  \bibfield  {author} {\bibinfo {author} {\bibfnamefont {D.~B.}\ \bibnamefont
  {Nelson}},\ }\bibfield  {title} {\bibinfo {title} {Conditional
  heteroskedasticity in asset returns: A new approach},\ }\href@noop {}
  {\bibfield  {journal} {\bibinfo  {journal} {Econometrica: Journal of the
  Econometric Society}\ ,\ \bibinfo {pages} {347}} (\bibinfo {year}
  {1991})}\BibitemShut {NoStop}%
\bibitem [{\citenamefont {Lee}\ and\ \citenamefont {Engle}(1999)}]{cGARCH}%
  \BibitemOpen
  \bibfield  {author} {\bibinfo {author} {\bibfnamefont {G.}~\bibnamefont
  {Lee}}\ and\ \bibinfo {author} {\bibfnamefont {R.}~\bibnamefont {Engle}},\
  }\bibfield  {title} {\bibinfo {title} {A permanent and transitory component
  model of stock return volatility},\ }\href@noop {} {\bibfield  {journal}
  {\bibinfo  {journal} {Cointegration, Causality and Forecasting: A Festschrift
  in Honor of Clive W.J. Granger}\ ,\ \bibinfo {pages} {475}} (\bibinfo {year}
  {1999})}\BibitemShut {NoStop}%
\bibitem [{\citenamefont {Zakoian}(1994)}]{TARCH}%
  \BibitemOpen
  \bibfield  {author} {\bibinfo {author} {\bibfnamefont {J.-M.}\ \bibnamefont
  {Zakoian}},\ }\bibfield  {title} {\bibinfo {title} {Threshold heteroskedastic
  models},\ }\href@noop {} {\bibfield  {journal} {\bibinfo  {journal} {Journal
  of Economic Dynamics and control}\ }\textbf {\bibinfo {volume} {18}},\
  \bibinfo {pages} {931} (\bibinfo {year} {1994})}\BibitemShut {NoStop}%
\bibitem [{\citenamefont {Glosten}\ \emph {et~al.}(1993)\citenamefont
  {Glosten}, \citenamefont {Jagannathan},\ and\ \citenamefont
  {Runkle}}]{glosten1993relation}%
  \BibitemOpen
  \bibfield  {author} {\bibinfo {author} {\bibfnamefont {L.~R.}\ \bibnamefont
  {Glosten}}, \bibinfo {author} {\bibfnamefont {R.}~\bibnamefont
  {Jagannathan}},\ and\ \bibinfo {author} {\bibfnamefont {D.~E.}\ \bibnamefont
  {Runkle}},\ }\bibfield  {title} {\bibinfo {title} {On the relation between
  the expected value and the volatility of the nominal excess return on
  stocks},\ }\href@noop {} {\bibfield  {journal} {\bibinfo  {journal} {The
  journal of finance}\ }\textbf {\bibinfo {volume} {48}},\ \bibinfo {pages}
  {1779} (\bibinfo {year} {1993})}\BibitemShut {NoStop}%
\bibitem [{\citenamefont {M{\"u}ller}\ \emph {et~al.}(1993)\citenamefont
  {M{\"u}ller}, \citenamefont {Dacorogna}, \citenamefont {Dav{\'e}},
  \citenamefont {Pictet}, \citenamefont {Olsen},\ and\ \citenamefont
  {Ward}}]{muller1993fractals}%
  \BibitemOpen
  \bibfield  {author} {\bibinfo {author} {\bibfnamefont {U.~A.}\ \bibnamefont
  {M{\"u}ller}}, \bibinfo {author} {\bibfnamefont {M.~M.}\ \bibnamefont
  {Dacorogna}}, \bibinfo {author} {\bibfnamefont {R.~D.}\ \bibnamefont
  {Dav{\'e}}}, \bibinfo {author} {\bibfnamefont {O.~V.}\ \bibnamefont
  {Pictet}}, \bibinfo {author} {\bibfnamefont {R.~B.}\ \bibnamefont {Olsen}},\
  and\ \bibinfo {author} {\bibfnamefont {J.~R.}\ \bibnamefont {Ward}},\
  }\bibfield  {title} {\bibinfo {title} {Fractals and intrinsic time: A
  challenge to econometricians},\ }\href@noop {} {\bibfield  {journal}
  {\bibinfo  {journal} {Unpublished manuscript, Olsen \& Associates,
  Z{\"u}rich}\ ,\ \bibinfo {pages} {130}} (\bibinfo {year} {1993})}\BibitemShut
  {NoStop}%
\bibitem [{\citenamefont {Corsi}\ \emph {et~al.}(2012)\citenamefont {Corsi},
  \citenamefont {Audrino},\ and\ \citenamefont {Ren{\'o}}}]{corsi2012har}%
  \BibitemOpen
  \bibfield  {author} {\bibinfo {author} {\bibfnamefont {F.}~\bibnamefont
  {Corsi}}, \bibinfo {author} {\bibfnamefont {F.}~\bibnamefont {Audrino}},\
  and\ \bibinfo {author} {\bibfnamefont {R.}~\bibnamefont {Ren{\'o}}},\
  }\bibfield  {title} {\bibinfo {title} {Har modeling for realized volatility
  forecasting},\ }\href@noop {} {\bibfield  {journal} {\bibinfo  {journal} {-}\
  } (\bibinfo {year} {2012})}\BibitemShut {NoStop}%
\bibitem [{\citenamefont {Panchal}\ \emph {et~al.}(2010)\citenamefont
  {Panchal}, \citenamefont {Ganatra}, \citenamefont {Kosta},\ and\
  \citenamefont {Panchal}}]{panchal2010searching}%
  \BibitemOpen
  \bibfield  {author} {\bibinfo {author} {\bibfnamefont {G.}~\bibnamefont
  {Panchal}}, \bibinfo {author} {\bibfnamefont {A.}~\bibnamefont {Ganatra}},
  \bibinfo {author} {\bibfnamefont {Y.}~\bibnamefont {Kosta}},\ and\ \bibinfo
  {author} {\bibfnamefont {D.}~\bibnamefont {Panchal}},\ }\bibfield  {title}
  {\bibinfo {title} {Searching most efficient neural network architecture using
  akaike’s information criterion (aic)},\ }\href@noop {} {\bibfield
  {journal} {\bibinfo  {journal} {International Journal of Computer
  Applications}\ }\textbf {\bibinfo {volume} {1}},\ \bibinfo {pages} {41}
  (\bibinfo {year} {2010})}\BibitemShut {NoStop}%
\bibitem [{\citenamefont {Karniadakis}\ \emph {et~al.}(2021)\citenamefont
  {Karniadakis}, \citenamefont {Kevrekidis}, \citenamefont {Lu}, \citenamefont
  {Perdikaris}, \citenamefont {Wang},\ and\ \citenamefont
  {Yang}}]{karniadakis2021physics}%
  \BibitemOpen
  \bibfield  {author} {\bibinfo {author} {\bibfnamefont {G.~E.}\ \bibnamefont
  {Karniadakis}}, \bibinfo {author} {\bibfnamefont {I.~G.}\ \bibnamefont
  {Kevrekidis}}, \bibinfo {author} {\bibfnamefont {L.}~\bibnamefont {Lu}},
  \bibinfo {author} {\bibfnamefont {P.}~\bibnamefont {Perdikaris}}, \bibinfo
  {author} {\bibfnamefont {S.}~\bibnamefont {Wang}},\ and\ \bibinfo {author}
  {\bibfnamefont {L.}~\bibnamefont {Yang}},\ }\bibfield  {title} {\bibinfo
  {title} {Physics-informed machine learning},\ }\href@noop {} {\bibfield
  {journal} {\bibinfo  {journal} {Nature Reviews Physics}\ }\textbf {\bibinfo
  {volume} {3}},\ \bibinfo {pages} {422} (\bibinfo {year} {2021})}\BibitemShut
  {NoStop}%
\end{thebibliography}%

\end{document}